# Re-Examining the Statistical Methodology and Onomastic Claims of Gregor and Blais' Argument from Name Popularity[*]

Jason Wilson, Ph.D. ORCID: 0009-0003-7734-1110

Department of Math and Computer Science, Biola University, La Mirada, CA, 90639, USA

jason.wilson@biola.edu

Luuk van de Weghe, Ph.D. ORCHID: 0000-0001-8710-503X

Independent scholar, Lenoir, NC, 28645, USA

luukvandeweghe@gmail.com

## Abstract

In 2024 Gregor and Blais published a *JSNT* article using two different statistical methods to conclude, contra Bauckham (2017), that selected Apocryphal texts and the Babylonian Talmud 'do not correspond to the distribution among first-century Palestinian Jews statistically significantly worse than the distribution in Gospels-Acts' and 'the two corpora paradoxically align better in some respects'. In this paper, we show that the first method is statistically invalid, and the second is the wrong tool for the job. This is in alignment with the critique of Van de Weghe and Wilson (2024) and in support of their use of the chi-squared goodness-of-fit test which established name occurrences in the Gospels and Acts, as opposed to Gregor and Blais' uniform, apocryphal, or Talmudic corpora, 'fit into their historical context at least as well as those in the works of Josephus'. Regarding this historical context, helpful insights are provided by Gregor and Blais regarding potential distortions within the onomastic reference distribution, and this article suggests a way forward, addressing orthographic issues, sample biases, several problems with the implementation of Gregor and Blais' inclusion criteria, and 87 new onomastic finds from ossuaries, ostraca, and documentary papyri that need to be incorporated into the lexicon. While legitimate concerns are raised by Gregor and Blais, several problems with their own onomastic datasets are also discussed.

# 1. Introduction

In 2002, Tal Ilan published the first volume of the definitive *Lexicon of Jewish Names in Late Antiquity*, which included the number of unique individuals with each name. In 2008 Richard Bauckham, based on his extensive study of the names of Palestinian Jews during the time of Jesus, observed that the most popular names in the Gospels-Acts tended to be the most popular names in Ilan's database. Similarly, the least popular names tended to be the least popular in Ilan's database. Bauckham's observation is displayed graphically in Figure 1. He updated the work in 2017. In 2023, Gregor and Blais published a critique of Bauckham's work. They employed the digitized version of Ilan's database and attempted a rigorous statistical analysis of Bauckham's claims. We commend them for those advances in the analysis of New Testament name frequency data. Unfortunately, there were some fundamental methodological flaws in their work, as shown by Van de Weghe and Wilson (2024). Despite this, in Gregor and Blais' next publication, *Re-examining Richard Bauckham's Argument from Name Popularity in Light of Apocryphal and Talmudic Evidence* (2024), they employed the identical flawed methodology.[1] In this paper, we will first argue along the lines of Van de Weghe and Wilson (2024) that the methodology used in Gregor and Blais (2024) is fundamentally flawed, and to the extent that their arguments are based upon their statistical model, they therefore do not hold. Next, we will critically interact with the onomastic data, discussing some helpful insights by Gregor and Blais, some new finds from the previous decade that need to be incorporated, and several onomastic observations that potentially call the rigour and perspective provided by Gregor and Blais into question.

---

[1] 'Before we proceed to test Bauckham's onomastic argument on the two extra-biblical corpora, it is necessary to briefly summarize Gregor and Blais's (2023b) methodology and relevant findings since we use their statistical model' (p. 281).

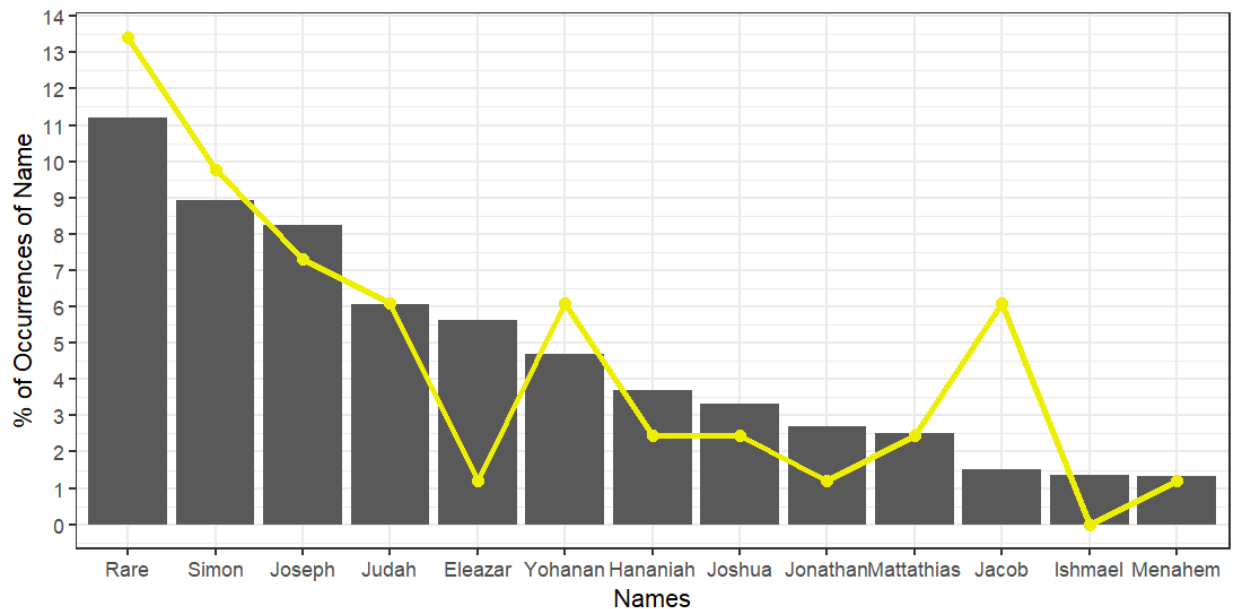

*Figure 1. Top 12 names' percentages. The grey bars represent the percent of total names from the historical reference (Ilan). The yellow line shows the same percentages from the full Gospels-Acts name list. The names shown are the 12 most popular male Jewish Palestinian names from the time of Jesus in the historical reference distribution, plus the tallest bar is the % of rare names (1 occurrence only).*

A word about the data is in order. Since our primary concern with Gregor and Blais' approach is methodological, it is not relevant at this point to describe the technical details of the datasets upon which the methods described in this paper have been used in Bauckham (2017), Gregor and Blais (2023, 2024) and Van de Weghe and Wilson (2024). The important thing to know is that all the methods considered use two different datasets: the first is a historical reference dataset drawn from all extant name references, which will be referred to as *Ilan,* and which will be discussed in more detail when we engage with the onomastic data later. The second is a test dataset from a particular text or combination of texts, e.g. Gospels-Acts, Apocrypha, and Babylonian Talmud. In Figure 1, the historical reference is the grey bars drawn from Ilan and the test data is the yellow line drawn from the Gospels-Acts. While there is some room for disagreement over the exact test datasets (Van de Weghe and Wilson, 2024), which sometimes amounts to a handful of names, the disagreements never change the conclusions of the first and primary method of Gregor and Blais. They do, however, change the conclusions of the secondary method, which will be discussed below. The debate is therefore about the methodology employed: Which method(s) are valid? Why?

In section 2 we will describe the methodological flaws of both the primary and secondary methods of Gregor and Blais (2024). In section 3 we revisit the Ilan reference distribution, engaging with some insights provided by Gregor and Blais, as well as additional concerns with both *Ilan* and some onomastic data provided by Gregor and Blais themselves. We conclude in section 4.

# 2. Critique of Gregor and Blais' statistical methodology

Gregor and Blais (2024) use two different statistical methods throughout their paper. In this section, we will describe their first method, critique it, then describe their second method, and critique it. The first method is the primary method and will therefore be our primary focus. The critiques are based on Van de Weghe and Wilson (2024). Each point of critique is distinct, although they bear different relations to one another. Our argument is a cumulative case argument: we contend that the added force of the separate points combine to make a strong case, even if some are discounted by the reader.

Although we assume that the reader is familiar with Gregor and Blais (2024), we hereby summarize their first statistical method for the sake of review and clarity. It is represented by Figure 2 (Gregor and Blais 2024: Fig. 1). The concept is to construct two candidate reference distributions to compare against one another: a historical distribution versus a uniform distribution, as summarized in the following three steps:

1. Historical distribution: '[E]very contested named character in Gospels-Acts existed in history'. This is represented in Figure 2. The names are in order, left to right, of their frequency in the Ilan database. A 95% confidence interval is constructed for each individual name separately on the basis of the Ilan's database.[2]
2. Uniform distribution: '[A]ll contested named characters, even entirely uncontroversial ones, were invented together with their names by tradents of oral tradition or literary authors who had absolutely no information about the popularity of Jewish names in first century Palestine'. This is the hypothesized completely random distribution, which will be referred to as the uniform distribution. The 95% confidence intervals are the gray shaded region, which consists of zero or one name occurrences.

---

[2] Gregor and Blais (2023) describe the historical distribution. In particular, they used the full Ilan distribution, minus the contested Gospels-Acts names, plus one 'artificial occurrence' of each Gospels+Acts occurrence for technical reasons.

3. If the test distribution is the Gospels-Acts names, the question is to see which reference distribution the Gospels-Acts fits better – the historical or the uniform?

4. Conclusion: 'Since 24 out of the 32 names in the contested Gospels-Acts sample are names with only one user in Gospels-Acts, the confidence intervals produced by the two opposite extreme scenarios are identical…. The two opposite extreme scenarios do not produce statistically significantly different results for the remaining eight names with more than one user in Gospels-Acts'. In other words, they conclude that their method is not statistically powerful enough to discriminate between the true historical distribution and the fictional uniform distribution. Therefore, 'Bauckham's thesis offers no advantage in explaining the observed correspondence between name popularity in Gospels-Acts and in the contemporary Palestinian Jewish population over an alternative model of "anonymous community transmission (Gregor and Blais 2023:1)"'.

Our first critique is that Gregor and Blais' conclusion does not follow from Figure 2.[3] We disagree with their interpretation and actually draw the opposite conclusion. The issue is whether the pattern of the data follows the historical distribution or a uniform distribution. These two shapes are displayed in Figure 3. It is obvious from the data, shown in both Figure 1 and Figure 2, that it fits historical distribution better than the uniform distribution. The confidence intervals themselves in Figure 2, even though wide, also match this pattern (notice the tops of the intervals). If Ilan approximates the true historical name distribution, then one in every twenty 95% confidence intervals would be expected to not contain the true name, on average. Of the 26 names shown in Figure 2, only one name (James) lies outside the historical Ilan confidence interval while seven names lie outside the uniform confidence intervals (the black diamonds).

---

[3] This is Figure 1 of Gregor and Blais (2023), which is reproduced as Figure 1 in Gregor and Blais (2024). However, for some unknown reason they dropped two important details: (i) the uniform confidence intervals (shaded gray region), and (ii) the colorization of the diamonds showing whether the Gospels-Acts name frequencies are inside both the historical and uniform confidence intervals (white), or just the historical distribution confidence intervals (black - with the one exception of James which has 3 occurrences, which is just above the 0-2 confidence interval).

Gregor and Blais' interpretation of their own figure is questionable.

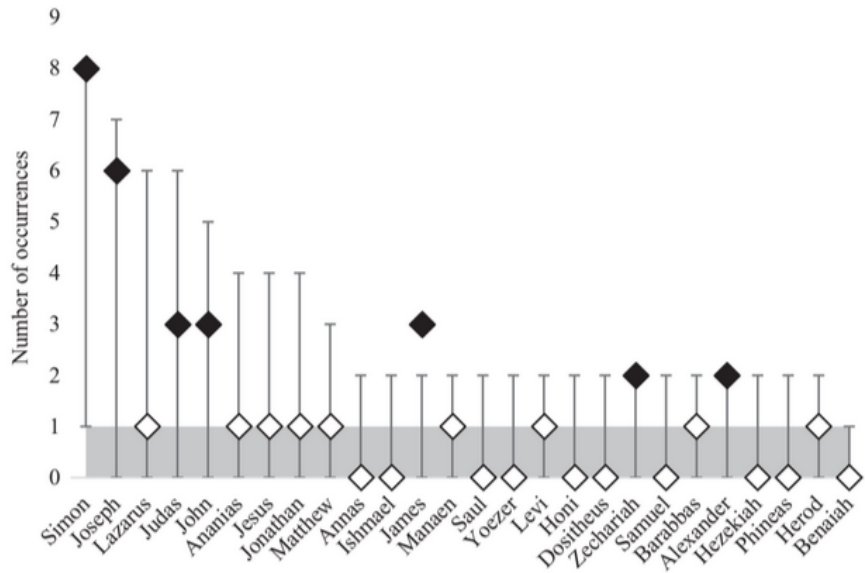

*Figure 2. Contested name occurrences in Gospels-Acts (diamond markers) with 95% confidence intervals calculated from the distribution of contemporary names. Black markers represent names with the number of occurrences outside the interval produced by random draws from a uniform distribution (coloured grey).*

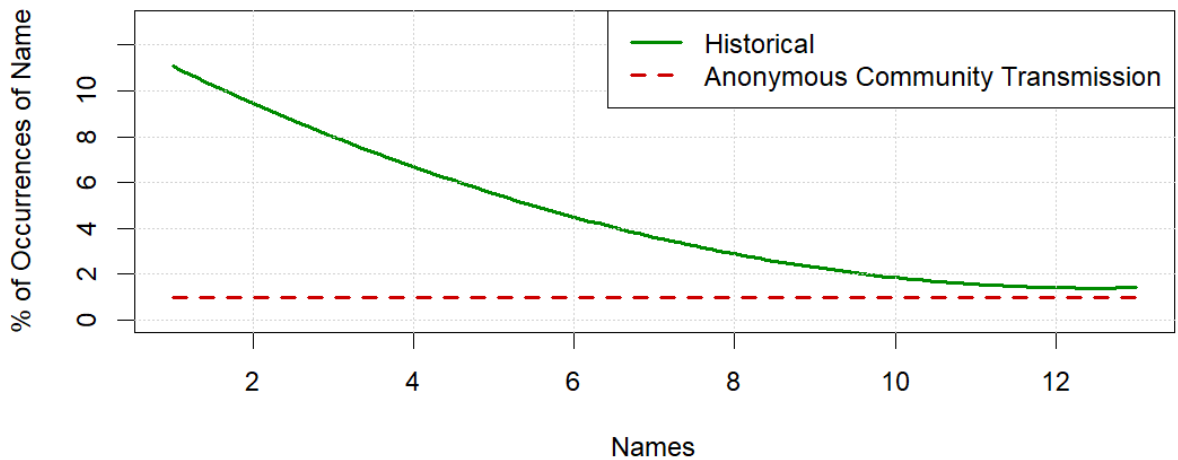

*Figure 3. Comparison of the historical and fictional distribution shapes. The historical distribution is characterized by the existence of popular and unpopular names, whereas the anonymous community transmission distribution has no popularity throughout.*

The second major critique is that Gregor and Blais implicitly assert a statistical hypothesis test with a null hypothesis of 'anonymous community transmission', or random name selection. The problem is they did not follow conventional statistical protocol and they reversed the null and alternative hypotheses. Indeed, standard statistical tests are available to formally test a candidate distribution against a reference distribution. In fact, the chi-squared test of goodness-of-fit is 'the most common and widely used statistical test of fit between a categorical test distribution and a categorical reference distribution (Van de Weghe and Wilson, 2024: 196)'. The reason the historical distribution must be the null hypothesis is that it is the only distribution which is known. If it fails, the alternative could be any other distribution (including, but not limited to, the uniform). The uniform distribution proposed by Gregor and Blais presents a theoretical extreme case; if it were the null hypothesis of a chi-squared goodness-of-fit test, then it would be rejected, which is the opposite of Gregor and Blais' conclusion.

Third, using the logic of Gregor and Blais, samples of size 53 from known historical texts would conclude 'fit' to the uniform distribution, i.e. 'anonymous community transmission'. To see this, look at Figure 2: the top of the grey bar (the uniform distribution 95% confidence interval) passes through, or at least touches, all the historical distribution confidence intervals shown. Because of this, Gregor and Blais argue that Gospels-Acts fit the uniform distribution. In their 2023 paper they show the same diagram, except using the writings of Josephus. The same pattern holds, except that the confidence intervals don't overlap – but only because Josephus has a larger sample size! If a subset of size 53 from Josephus were used, or a different known historical author with that sample size, the confidence intervals would correspondingly widen and look like Figure 2. This is because the mathematical formula for the width of confidence intervals includes $\frac{1}{\sqrt{sample\ size}}$, which shrinks as the sample size increases. Following the logic of Gregor and Blais' method, they would be forced to conclude the writings of Josephus fit the uniform distribution and therefore the names were plausibly fabricated. However, scholars generally do not think Josephus' writings were generated by anonymous community transmission. Such a conclusion would be a result of the flawed methodology. Using the proper method, historical samples demonstrate a fit with the historical reference population regardless of their size and are statistically distinguishable from Gregor and Blais' uniform or mixed distributions at sizes comparable to the Gospels-Acts sample. We revisit the issue of sample size in Section 3.2.2.

The fourth critique is that the confidence interval method used by Gregor and Blais (Figure 2) falsely assumes independence between names. Confidence intervals are regularly used in this way because they make a visualization which helps explain the point. Nevertheless, such multiple univariate confidence interval figures should not be used drawing final conclusions, but rather just for the visualization. The proper next step is a formal statistical test, with a multiple testing adjustment (Gelman and Hill 2007: 452-455). The reason is that the proportion of one name is dependent upon the proportions of all of the other names, which is not accounted for in the confidence interval calculations. For example, 8.9% of male Jews were named 'Simon' in Jesus' day; if the popularity of the name 'Simon' dropped in half (8.9/2 = 4.45), then the total percentage of the other names would increase by 4.45%. Now, this dependence relationship is precisely accounted for by the chi-squared goodness-of-fit test used by Van de Weghe and Wilson (2024).

The fifth critique is that Gregor and Blais assert that names occurring only once in Gospels-Acts (e.g. *rare* names) provide no information about whether Gospels-Acts fit the historical distribution. 'This means that almost the entirety of Bauckham's sample is wholly uninformative, no different from "white noise" produced by random chance'.[4] However, grouping individuals in order to meet the test conditions is a common issue in goodness-of-fit testing.[5] The information for every name can be accounted for by binning them into appropriate groups. For example, in Figure 1, the first bar represents the 11.2% of rare names in the Ilan database, as compared with the 13.4% of rare names in Gospels-Acts, which is pretty close and appears to favour of Bauckham's thesis.[6]

Next, let us consider the second method employed by Gregor and Blais: using a probability method to separately analyse individual names or clusters of names.[7] Here are the steps:

---

[4] Gregor and Blais 2024, p. 282.
[5] For example, it is standard when analysing race/ethnicity statistics to use the most common race/ethnicity groups, and bin the remainder into an 'other' category. As another example, when considering waiting times, responses may be grouped as 1, 2, 3, 4, 5, 6, 7+, where the last category includes all waits of 7 or more days. These are characteristic examples and both are considered in the open source Diez, Barr, & Çetinkaya-Rundel (2022: Sec 6.3).
[6] Note that this 13.4% assumes the inclusion of the Acts 6.5 Hellenists (minus Nicolas from Antioch), which Gregor and Blais acknowledge should likely be included, contra Bauckham, but in line with Ilan, Van de Weghe, and Wilson (see Gregor and Blais, 2023: 185). They do not, however, incorporate them into their dataset. See our discussion on rare names below. Note also that this percentage (13.4%) has been calculated using our revised Ilan-1 dataset (see Section 3.1 below).
[7] Technically, the method is equivalent to a formal binomial hypothesis test. Enough information is not provided in Gregor and Blais (2024) to tell, but in Gregor and Blais (2023) it appears to be a randomisation test version, which is equivalent.

1. Let X be the number of characters with a particular name recorded during the time period. Then Gregor and Blais construct figures, called histograms, showing probability of observing different values of X, such as 0,1,2,3,4,5,6,7,8,9,10.

2. The probabilities in the histograms critically depend upon the assumptions made.

3. For the case X = number of Simons, if we were to randomly draw 53 names from the historical distribution,[8] 'the probability of drawing eight occurrences of the name Simon, the value observed in Gospels-acts, or more, … is only 2.9%'. Therefore, they conclude, Simon is 'overrepresented'.

4. For the case X = 'number of rare names in the sample of contested Gospels–Acts characters', Gregor and Blais (2023) show a histogram. They count 4 rare names in contested Gospels-Acts.[9] '[T]he probability of drawing four rare names… is relatively very small'. They don't provide the number, but using their assumptions we calculate 1.1%, which is indeed relatively small.

Our first critique of their second method is that they are simply inconsistent. We previously cited their writing that the rare names are 'wholly uninformative'. Gregor and Blais (2023) wrote that rare names were 'without any information about name popularity'. Only one of the two propositions can be true: rare names either do offer information about name popularity, or they don't, but not both. We affirm that they do offer information, as previously stated, and therefore welcome Gregor and Blais' expansion along these lines.

The second critique of the second method is that they reversed the null and alternative hypotheses from the first method. Although we agree that this is now the correct configuration of the hypotheses, it is a second inconsistency.

The third critique is that the methodology employed is highly sensitive to the assumptions made. In the case of Simon, using their probability method they reported a 2.9% probability of getting 'eight Simons… or more'. However, using the proportion of Simons in the historical data, $184/2185 = 8.4\%$, the theoretical calculation gives a 7.4% chance of randomly selecting 8 or more Simons in a random sample of 53 people.[10] However, this is biased because it considers only contested figures. Using the actual 82 figures in Gospels-Acts, there is a 38.6% chance of randomly selecting 8 or more Simons. Thus, depending on the assumptions, the probabilities range from 2.9% to 7.4% to 38.6%, which goes from unlikely to quite plausible.

---

[8] There are 82 Palestinian Jewish men in the Gospels+Acts, but Gregor and Blais argue for only considering the "contested" number, because everyone agrees the uncontested men are historical (2023: 190).
[9] Gregor and Blais (2023: 298) and Gregor and Blais (2024: 291).
[10] Since X = number of Simons, using the binomial distribution,

The fourth critique is that the methodology is inappropriate for the purpose. It was used to draw conclusions about isolated cases of names, or clusters of names, from a list of dozens of names. What criterion was used to select those few cases? Such isolated analyses are subject to the *multiple testing problem,* which is the inflation of the false positive error rate in the face of many tests on the same data set. This problem does not appear to have been recognized, let alone addressed.[11]

Moving to the argument from the 'rare names in the sample of contested Gospels-Acts characters', Van de Weghe and Wilson (2024) used similar reasoning to address it:

> *[Gregor and Blais'] conclusion … is highly sensitive to the number of rare names in the list, and we dispute their list, which stands at four (Aeneas, Agabus, Bartimaeus, and Timaeus). In their handling of the data, Gregor and Blais removed 26 "attested" name occurrences (21 names) from Bauckham's original 79 (45 names) to obtain their 53 (32 names) unattested occurrences. The names removed include two rare names (Qaifa and Toma). According to their calculations, the "probability of getting [four or fewer] names … is only 1.1%!"  According to GB's argument, the table below shows the number of rare names and the probability of that many or fewer occurring, if the Ilan-1 distribution was true. It can be seen that their conclusion is extremely sensitive to the number of rare names. GB's list gives a 1.1% probability, which would jump to 7.3% if just the two names were not excluded (Table 4). Moreover, if the Palestinian Acts deacons were added, as we believe they should be, it includes four more "rare" names and would bring the list to eight or ten rare names at 24% or 50% of occurring (Table 4), depending on whether we also add the two contested rare names, Qaifa and Toma.*

$$P(X \geq 8) = 1 - \sum_{x=0}^{7} \frac{53!}{x!\,(53-x)!}(.084)^x\,(.916)^{53-x} = 0.074$$

For more on these calculations, please see the Supplementary Materials listed in the references under our names.

[11] The multiple testing problem is encountered and addressed by Van de Weghe and Wilson (2024: 198).

| # rare | 3 | 4 | 5 | 6 | 7 | 8 | 9 | 10 |
|--------|-----|-----|-----|-----|----|----|----|----|
| % | 0.3 | 1.1 | 3.2 | 7.3 | 14 | 24 | 36 | 50 |

*Table 4. Probability of obtaining numbers of rare names. This table is based off Gregor and Blais (2023: Fig. 3). For example, # rare = 6 and %=7.3 means that there is a 7.3% probability that 6 or fewer rare names would be obtained in a sample of size 53 selected randomly from Gregor and Blais's pool of 2,582 occurrences.*

The point of this critique is to show how sensitive the second methodology is to the assumptions, which are themselves disputable.

In light of the above problems, is there a valid statistical method that can be used to discriminate between the ancient manuscript name lists which match the historical reference from those which fail to match it? The answer is *yes*. Van de Weghe and Wilson employed the standard statistical hypothesis test for this purpose called the *chi-squared test of goodness-of-fit test.* Other valid statistical methods for discrete probability distributions might also be employed, such as a Bayesian analog (Gelman, Meng, and Stern 1992) or a randomisation method. The problem with Gregor and Blais' first method is that it is not valid, in the sense that it would not reliably obtain the truth if used repeatedly in similar contexts. The second method is valid, but its use is only for specific cases and not capable of reaching a conclusion about Bauckham's thesis, in general. Furthermore, the second method is highly sensitive to assumptions, and the stated assumptions were shown to be dubious. By contrast, Van de Weghe and Wilson's use of an appropriate statistical method, the chi-squared goodness-of-fit test, led them to conclude the Gospels-Acts 'accurately retained personal names – those unmemorable pieces of personal information – at a remarkably high level' (Van de Weghe and Wilson 2024: 214). Goodness-of-fit testing, however, demonstrates that Gregor and Blais' uniform distribution (Gregor and Blais 2023) does not reflect the name population of Jesus' Palestine, and neither do their additional apocryphal and Talmudic corpora (Gregor and Blais 2024), all of which runs directly contrary to the stated conclusions of their recent article.[12] We will discuss these corpora in more detail in Sections 3.2.3 and 3.2.4.

---

[12] The p-value of their uniform sample is $1.69 \times 10^{-15}$ (Van de Weghe and Wilson 2023: 200); the p-values of their Talmud and apocryphal samples are .00263 and .00004, respectively. For the definition of p-values, see Section 3.1 below. Gregor and Blais state: 'Using the most recent onomastic data and substantively enhanced statistical analysis,

Regarding statistical methodology, we rest our case and recommend that readers disregard any conclusions which follow from the flawed graphical method of Gregor and Blais (2024) and be very cautious with the probability method due to its inappropriateness for the job and sensitivity to the assumptions. That said, in the process of drawing their conclusions, Gregor and Blais scrutinize the onomastic data carefully and capably interact with Bauckham on a level which does not depend upon their statistical methods. In the next section, we shift to interacting with Gregor and Blais (2024) on this level.

# 3. Onomastic Data

## 3.1 Ilan's Reference Population

Before addressing some potential deficiencies with Gregor and Blais' onomastic data, we turn to legitimate concerns they raise regarding Ilan's reference distribution and the problem of nonrandomness within the onomastic datasets. Gregor and Blais suggest that disproportionate ossuary findings and historical figure references are attributed to the Jerusalem area, which could incline the names in the Gospels-Acts to more favourably reflect *Ilan* for the following reason: 'The letters of Paul illustrate that even prior to the composition of the Gospels, correspondence and personal visits exposed members of Christian communities, including those outside Palestine, not only to specific Palestinian Jewish names … but specifically to names of Jewish Christians active in Jerusalem. … [T]his exposure might have led to a tendency to supply invented characters with Palestinian names popular specifically in the Jerusalem area' (Gregor and Blais 2024: 277).

While it is unclear how many of these 'Jewish Christians active in Jerusalem' were in fact native to Jerusalem and not, say, to Galilee or other regions frequented by Jesus, the overall observation concerning the Jerusalem provenance of most ossuary inscriptions is correct. Gregor and Blais also note the over-representation of certain Greek names in Josephus' writings which, we show that even in textual corpora containing large numbers of recognizably fictitious characters, name probability distributions do not align statistically significantly worse to the distribution among first-century Palestinian Jews' (2024: 293).

again, is a valid observation (Gregor and Blais 2024: 277–78). But how might these impact the Ilan reference distribution and its relationship to the Gospels-Acts names? For it to make a difference, there would have to be a substantive difference in the Jewish name distribution of the Jerusalem area compared to the rest of Palestine.

One way to assess this significance would be to test the relevant data samples against various iterations of *Ilan*: for example, one iteration of ossuary finds; one of ostraca, graffiti, numismatic, and papyri finds; and one of literary name occurrences. Each of these datasets contains distinct potential sample biases. Whereas ossuary inscriptions are confined mostly to the Jerusalem vicinity, they are easily datable to before the end of the Jewish War and identified as Jewish; other inscriptional data, while being more geographically diverse, can present further difficulties in establishing firm dating parameters and Jewish nationality; literary works provide more context in this regard for dating, provenance, and Jewish nationality, yet these texts often favour an elite population and Greek versus Semitic name attributions. Fortunately, if we separate *Ilan* into these three iterations (ossuaries, other inscriptions+papyri, and literary), Gregor and Blais' own Gospels-Acts sample has very similar p-values in relation to all three: 0.3093; 0.5753; 0.2701. The p-value is a technical statistic.[13] It is the probability that the set of names observed, or more extreme, would be observed if they were randomly selected from the *Ilan* reference set. The usual benchmark for comparing p-values is .05. Since the Gospels and Acts p-values are well above .05 in each of these cases, the Gospels and Acts are considered to match the Ilan reference distribution, even when broken into these iterations. By 'match' we do not mean their fit to the historical distribution is mathematically proven. Rather, it means the test indicates that the name frequencies from the sample are within what would be expected by chance from the Ilan reference, with that sample size.

For now, *Ilan* is the best historical reference we have, and we note, in agreement with Gregor and Blais, that any argument hinges upon it representing the true name distribution.[14] That said, we must commit ourselves to finding and using the most accurate data available to us, and to take all reasonable means to ensure that our data best reflects the historical situation we are examining. It has come to our attention that Ilan's database, even in Hayim Lapin's machine-readable format, is outdated.[15] Furthermore, when Gregor and Blais scraped the Ilan data with their inclusion criteria (occurrences had to be personal names of Jewish males from Palestine, not definitely datable outside 4 BCE–73 CE; criteria also adopted by Van de Weghe and Wilson, 2024), they did not conduct contextual evaluations of Ilan's references and source materials to more carefully determine eligibility.

---

[13] Discussed in detail by Van de Weghe and Wilson (2024: 196–98).

[14] This is in line with the comments made by Gregor and Blais (2024: 278): 'nothing can be done to compensate for systematic biases present in Ilan's data since their statistical properties are unknown. We are, therefore, forced to treat the data as if it were a truly random sample of all ancient Jewish names'.

This process would result in the removal of at least 73 additional occurrences, many for contextual reasons.[16] Here we give but a few samples of occurrences that should be removed. Sosipatrus (Σωσίπατρος, AJ 14.241) was an ambassador of Hyrcanus the high priest, well before the 4 BCE–73 CE timeframe. Many other name occurrences, like Judah ben Tabbaa (active c. 80 BCE), Nittai the Arbelite (one of the second 'pairs' of mAb1, living c. 120 BCE) and Joseph ben Yoezer, a famous rabbi from the early Maccabean period, are outside of Gregor and Blais' specified timeframe; Ilan's date ranges, at times quite broadly applied, are sometimes imprecise. Several occurrences, like Ezra (CJO, no. 822; cf. CIIP 519), result from incorrect interpretations of inscriptional markings (in this case, 'Avira' being the more likely reading). Ostraca inscriptions, like that of a certain חנניה (Naveh 2000: no. 9; cf. CIIP 610), are again dated broadly by Ilan to 'pre-70 CE' while more careful analysis renders the inscription outside of the timeframe: in this case, as a likely product of the 4th c. BCE.[17]

Name occurrences from recent CIIP volumes and other publications should also be incorporated into the reference population. Ostraca findings from Herodium, Qumran, and Machaerus, for example, should be included.[18] Additional ossuary inscriptions have also been catalogued from Shu'afat and Qatamon, as well as papyri from the Judean Desert.[19] These are but a few samples from the 87 new occurrences we were able to introduce into *Ilan*.[20] While statistical methodology must assume randomness among the test and reference samples – and this can only be tentatively assumed for onomastic datasets – the above exercise of separating *Ilan* into various iterations has demonstrated that serious scepticism toward the legitimacy of the data is not warranted. Even Josephus' emphasis on Greek names was, for example, statistically demonstrable, as noted by both Gregor and Blais (2024) and Van de Weghe and Wilson (2024). Considerable nonrandomness was introduced into the test samples, however, by Gregor and Blais when they selected only name occurrences under the criterion that these needed to be otherwise unattested in historical records (Gregor and Blais 2023, 2024). Unlike criteria imposed by Bauckham, Ilan, or Van de Weghe and Wilson in other studies – the aims of which were to allow the most natural presentation of original name statistics and distributions – this imposition by Gregor and Blais is artificial and thereby introduces an additional level of sample bias.[21] Yet even with this additional introduction of bias, Gregor and Blais' version of the Gospels-Acts sample shows a consistent fit with *Ilan* in all three of its iterations, as discussed above.

---

[15] Lapin's database is the machine-readable form of Tal Ilan's lexicons; it has been utilised in all formal statistical analyses on the Gospels-Acts to date. It is found here: https://github.com/hlapin/eRabbinica/tree/master/ilanNames (accessed 16 May, 2025).

[16] The datasheet titled 'Removed Names' in our Supplementary Materials gives the precise reference and justification for the removal of each of these 73 name occurrences.

[17] This is not to denigrate the available data that we have which, in the vast majority of cases, is well-referenced and confirmed to be accurate. Having been in correspondence with Tal Ilan about her lexicon, we concur that the minor errors, sample biases, and the relative incompleteness of her database would not change any of the statistical conclusions that have been reached. But it seemed incumbent on us to move beyond the mere impression of this fact and confirm it with further evaluation. Van de Weghe and Wilson's scraped data (2024) should eventually also be cross-referenced with the data scraped by Gregor and Blais (2023) to highlight and adjust for errors in scraping. This is a current area of work.

[18] Yardeni (2013): 209-43; Misgav (2013): 259-76; Eshel (2015): 460-73.

## 3.2 Gregor and Blais' Onomastic Data

While Gregor and Blais do an admirable job of drawing our attention to some concerns regarding *Ilan* and make useful observations about minor errors in Richard Bauckham's *Jesus and the Eyewitnesses* (2017), a few corrections are warranted.

---

[19] Cf. Adawi, Nagar Katsnelson (2013): no. 1-9; CIIP 1/2, no. 1088.

[20] We are very grateful to Tal Ilan for sharing with us about 2000 onomastic finds from the last decade and beyond (since the publication of her 2012 addendum to Vol. 1) in order to sift through these and include the relevant contributions. A more extensive discussion of the improvements awaits a future publication, but the revised version of Ilan, titled 'Revised Ilan', is available in our Supplementary Materials. The p-values for the three iterations are drawn from this updated reference distribution. Gregor and Blais' earlier version of *Ilan*, which was also used by Van de Weghe and Wilson (2024), had several names missing (e.g. Malka and Sheila), which is easy to do, due to the difficulty of scraping the Ilan database. Our Revised Ilan is adapted from the data scraped by Van de Weghe and Wilson (2024) due to the lack of reference material available in Gregor and Blais' original dataset.

[21] Note Ilan's comment on p. 2 (Ilan 2002): 'The most important results that can be arrived at from such a lexicon [as hers] are of a statistical nature…[since] this is a large corpus, it is assumed, on the basis of the statistical theory of probability that such a record adequately demonstrates the patterns of name-giving that prevailed among Greco-Roman Palestinian Jews'.

### 3.2.1 Name Designations

Gregor and Blais (2024: 280) state that Bauckham 'inexplicably omits Chuza, a steward of Herod Antipas in Lk. 8.3' from his list of Jewish name occurrences. But this is not inexplicable. Chuza is likely Nabatean and should therefore not be counted as a Jewish name occurrence, a position aptly defended by Richard Bauckham himself in a lengthy discussion covering the Nabatean inscriptional attestations of this name (Bauckham 2002: 361–77). Gregor and Blais (2024: 284) also claim to follow Ilan's orthography and that Elisaeus, from a *Clementine Homilies* list, represents the same name as Eleazar. Yet Elisaeus is represented as *Elisha* under Ilan's orthography (Ilan 2002: 63), with her reference notes clearly linking this occurrence to the *Clementine Homilies*.[22]

Sometimes apparent counting errors also encroach on otherwise careful scholarship. Gregor and Blais state, for example, that their apocryphal corpus contains 50 fictional name occurrences, but in their article they mention and/or list 51 while there appear to be, according to our data, 52 (14 for *Clementine Homilies*; 25 for *Acts of Pilate* (not 24); 13 for *Book of the Bee*; see Gregor and Blais 2024: 285–86). Like the inconsistencies that they highlight from Bauckham's work, these minor errors or discrepancies are fortunately inconsequential. It is our informed conclusion that the proper statistical method will, and does, show a consistent improvement in the Gospels and Acts data over apocryphal samples whether one uses Gregor and Blais' dataset, Bauckham's dataset, or the dataset recently presented by Van de Weghe and Wilson (2024).[23] More on this below.

---

[22] Regarding orthography, Van de Weghe and Wilson also incorrectly identified several names from *Ben Hur* due to orthographic variants, and contrary to their published claim (2024: 205), the revision of this data causes this fictional novel to pass the goodness of fit test (p-value .1113). More on this below. Problems of orthography are another feature of Ilan's lexicon itself, which involves interpretations of how to classify names that can be questionable. Hananiah and Hanan, as well as Honi, stem from the same Semitic root, חנן, to show favour + יה: 'Yahweh has shown favour', and yet Ilan considers them distinct names based on functional versus morphological factors (Ilan 2002: 100). yet Ilan considers Eleazar (אלעזר) and Eliezer (אליעזר) under the same name, while epigraphical and papyrological sources apparently separate them; Josephus, however, and several rabbinic citations may appear to conflate them. We thank Richard Bauckham for these insights, and for Tal Ilan for providing a further justification of her choices to designate certain variants and hypocoristics under specific English designations.

[23] The p-values for the name frequences of these samples are cited by Van de Weghe and Wilson (2024: 190, 200); Bauckham's list has a p-value of .6106; Gregor and Blais' list of 'contested names' has a p-value of .2749; Van de Weghe and Wilson's list has a p-value of .8556 (later updated to .7939 in v3 of their data).

## 3.2.2 Onomastic Congruence

Gregor and Blais (2024: 276) correctly highlight the shortcomings of Van de Weghe's *NTS* article (2023) in that it did not provide a robust statistical analysis of names.[24] Yet Gregor and Blais potentially miss some essential context that Van de Weghe's argument provides. Van de Weghe argued that a feature termed 'onomastic congruence' was present in the Gospels and in other historiographic biographies from the Early Empire, seeing it as an unintentional byproduct of authentic information-gathering and data retention; he defined onomastic congruence as follows: '1) a relatively significant number of appropriate proper names; 2) a relatively increased level of detail in proper names; 3) patterns of proper names reflecting "the situation on the ground"'(Van de Weghe 2023: 97). While due attention has been brought to Item 1, 'a significant number of appropriate proper names', Item 2 and Item 3 are also pertinent.

Regarding Item 2, the issue of onomastic detail (i.e. sample size) does appear to influence the male names samples that have been discussed throughout prior studies. Van de Weghe and Wilson, for example, collected the onomastic data for the historical novels, *The Spear* and *Ben Hur*. As a single corpus, these works have a low p-value, .0000000000000001; but contra Van de Weghe and Wilson, *Ben Hur* by itself, with only 31 names, is not statistically distinguishable from a historical text (p-value .1113; see footnote 21 above); likewise, while Gregor and Blais' combined apocryphal sample has a very low p-value (.00004), the *Clementine Homilies* by itself, with only 20 names, has a p-value of .4139, well above the standard threshold. Van de Weghe suggests that for a corpus or composition to contain enough onomastic detail, it typically needs 'around forty names' (2023: 97). At samples above this size, datasets from historiographical narratives consistently have high p-values, while samples from fictitious narratives have values below the standard threshold. For Jewish male names at least, this second aspect of onomastic congruence is corroborated by the onomastic evidence from the datasets analysed up to this point.

More importantly, Gregor and Blais do not consider a further detail of onomastic congruence: Item 3, patterns of proper names reflecting 'the situation on the ground', i.e. the appropriate distribution of qualifiers. Qualifiers often serve to disambiguate, to *distinguish* a person from someone else that might have the same name (e.g., John 'son of Zebedee'). Not all qualifiers disambiguate; sometimes they function as a title (e.g. 'Caiaphas *the high priest*'); for this reason, titles such as 'ruler', 'tetrarch', 'high priest', or 'king' are excluded from Table 1 below, which displays the number of disambiguating qualifiers in the Gospels-Acts and in the corpora analysed throughout all prior studies in this area.[25]

---

[24] Note that we updated the Table 1 from this paper in v3 of our Supplementary Materials (Van de Weghe and Wilson: 2024). We are currently involved in re-appraising the entire study, using better data-gathering and methodology, including utilising an improved version of Ilan-1, scraping of the DPRR within the appropriate parameters, and re-cataloguing the Greek names from relevant compositions in their lexical forms.

| *Qualifier Occurrences* | Matt. | Mark | Luke-Acts | John | Acts of Pilate | Book of Bee | Clem. Hom. | Ben Hur | The Spear |
|---|---|---|---|---|---|---|---|---|---|
| *Total Qualifiers* | 15 | 13 | 15 | 9 | 10 | 28 | 14 | 11 | 21 |
| *Male Names 102+ occur.[26]* | 9 | 7 | 9 | 6 | 3 | 10 | 2 | 4 | 9 |
| *Male Names 6-101 occur.* | 5 | 5 | 5 | 2 | 5 | 8 | 6 | 4 | 3 |
| *Male Names <6 occurrences* | 1 | 1 | 1 | 1 | 2 | 10 | 6 | 3 | 9 |

*Table 1. Disambiguating Qualifiers for Male Names, with Reference to Ilan*

In Table 1, the first row gives us the total number of males whose names contain a disambiguating qualifier. The second through fourth rows give us the number of names in the text with qualifiers, binned into three sections from the most popular to the least popular. For example, the Gospel of Matthew has 15 total names qualified; 9 out of 15 belong to men having one of the top five most popular names (those having 102+ occurrences in Ilan: Simon, Joseph, Eleazar, Judah, or John); 5 of 15 belong to males with standard common names (Jesus x2,

---

[25] For our full data on name qualifiers, see our Supplementary Materials. Qualifiers from Josephus and the Babylonian Talmud could not be included due to the difficulty involved in sifting through that amount of data. Female name qualifiers were also tallied but not included in the table because prior studies have focused exclusively on male names, due to the prominence in antiquity, making them more suitable for statistical analysis. A study devoted exclusively to female name occurrences is, however, an area of current research.

[26] After the list of the five 102+ occurrence names, Hananiah is in sixth place with 75 occurrences, showing a considerable step down. For <6 occurrences, this number was derived from the least popular name bins in Van de Weghe and Wilson (2024: 202, Figure 2). When determining whether a qualified name occurred less than six times in Ilan-1, historical occurrences from that text were not included, to create independence.

Matthew, and James x2), and only 1 of 15 qualifiers in Matthew belongs to a male with an unpopular name (in this case, to Andrew). This is expected under an authentic name-retention scenario, wherein persons with the most popular names need to be disambiguated much more frequently in a historical setting relative to those with less popular names, which would not need to be qualified for the purpose of disambiguation. The pattern of disambiguating popular names much more frequently than unpopular and standard names is less consistent in male samples from comparable fictional texts.

### 3.2.3 The Talmudic Data

This brings us to a further contrast with Van de Weghe's initial study. Van de Weghe contextualized his comparative study by gathering his names from narratives 'which were *firstly comparable in length to the Gospels-Acts* and secondly *centred around one or two protagonists*' (2023: 97; emphasis ours). In this regard, the Babylonian Talmud, despite Gregor and Blais' contention (2024: 287), does not represent a good parallel for what a fictionalizing author of the Gospel tradition could or would accomplish particularly because it is neither a historical nor a fictional narrative, but a multi-generational, redacted anthology of rabbinic debate and tradition. Names in the Talmud that may be fictional were formulated within a rabbinic context shaped by the halakhic and exegetical mesorah transmitted through named rabbis across many generations – particularly those found in later midrashic and Amoraic sources, which account for many of the Babylonian Talmud entries included in Tal Ilan's lexicon. Nevertheless, Gregor and Blais' corpus from the Babylonian Talmud, with 87 name occurrences (18 fictional), has a very low p-value: .00263.[27] Contrary to Gregor and Blais' claims, the proper statistical test is able to distinguish between comparable corpora of mixed historical/fictional samples (which do not accurately reflect the population distribution of Ilan-1), versus the Gospels-Acts sample (which plausibly reflects the population distribution). This trend continues with the apocryphal data.

---

[27] We replicate it in the Data and Supplementary Materials. Note that Pandeion (bBer 55b; Ilan 2002: 450) was not uploaded into the digitized version of Ilan's lexicon and was added to the list. Both with and without Pandeion, Gregor and Blais' sample has a very low p-value (w/ Pandeion: .004911), and this is also true if we run the Talmud sample against our own revised version of *Ilan* (p-value: .0001853).

## 3.2.4 The Apocryphal Data

The appropriate statistical method demonstrates that the Gospels-Acts name occurrences fit within a population represented by Ilan, but Gregor and Blais' apocryphal works do not. As stated earlier, the Gospels and Acts names fit whether we use Bauckham's list of names, Van de Weghe and Wilson's list, or Gregor and Blais' list; using the goodness-of-fit test, these have p-values of .6106, .2749, and .7939 respectively, showing a match to the Ilan reference distribution. Conversely, since the p-value for the Apocryphal names is .00004, which is well below .05, it is not considered to match the Ilan reference distribution.[28] See Van de Weghe and Wilson (2024: 196–98) for further technical details. This runs directly contrary to Gregor and Blais's stated conclusion.[29]

What remains unstated, however, is how deeply problematic Gregor and Blais's data from the *Book of the Bee* is within their corpus. The vast majority of 'contested' Gospels-Acts names in Gregor and Blais' apocryphal corpus come from this one composition – we were only able to count five within *Clementine Homilies* and *Acts of Pilate* (Nicodemus, Joseph of Nazareth, Lazarus, Barabbas, and Joseph of Arimathea). Even if there is a small discrepancy of several names, this means that Gregor and Blais likely scraped around 25 to 30 Gospel name occurrences from the *Book of the Bee*. This is how they can show seven Simons, six Josephs, and four Eleazars (although this should be three, since they incorrectly count Elisaeus) in their Figure 3 (2024: 286). These numbers are close to historically expected because many come from Gospel characters listed in the *Book of the Bee*. What is not included, however, are the many names of possible, even *likely*, Palestinian Jewish characters from that text that are not considered by

---

[28] Gregor and Blais' dataset was not readily available. We have aimed to reconstruct the list to the best of our ability based upon the lists and comments they provided (2024: 284–285); it is available in our Supplementary Materials. We have also provided complete name lists for all these compositions. While Gregor and Blais state that *Book of the Bee* has 13 fictional occurrences from Ilan-1, this is only true if 'Adi' is left in, which we did. Regardless of whether we should deem 'Adi' as historical, this person would remain contested (i.e., not externally attested in Ilan), so it is surprising that Gregor and Blais removed the name in any case (see their discussion in Gregor and Blais 2024: 285).

[29] For example, they state that 'even in textual corpora containing large numbers of recognizably fictitious characters, name probability distributions do not align statistically significantly worse [than the Gospels-Acts] to the distribution among first-century Palestinian Jews' (Gregor and Blais: 2024, 293); again, on p. 286 they state that 'not even 50 out of 82 contested occurrences being fictitious is sufficient to "skew" the name popularity distribution in the corpus enough to produce results that would be empirically distinguishable from what we see in Gospels-Acts'.

Gregor and Blais in their list; these are clearly fictitious occurrences that, when considered, demonstrate the overtly unhistorical nature of this composition. Distinguishing between historical characters, fictional characters with historical Jewish names, and allegedly historical Jewish characters with non-Jewish names becomes, to say the least, deeply problematic. Indeed, attempting to derive Gregor and Blais' 'contested' Gospels-Acts name occurrences from within this composition became quite difficult, so that we had to consult the statistics from their Figure 3 to recreate their corpus (in our case, this created 83 versus their 82 name occurrences in total).[30] To clarify the problem, let us discuss some onomastic features from the *Book of the Bee*. Here we will reference the same edition as that used by Ilan and, presumably, by Gregor and Blais (Budge 1886). On p. 112, one of the Twelve is called 'Labbaeus, who is surnamed Thaddaeus'. According to p. 106, Labbaeus is the same person as 'Jude the son of James'. The name occurrence, 'Labbaeus', not in Gregor and Blais' list. Yet this is a clear case of a Palestinian and Jewish name occurrence. In the list of the 70, we have names that give some indication that the author is trying to recreate Jewish names, with Bar Kubba, Addai, and Aggai being placed alongside names like Simon, Joseph, 'another Judah', Levi, and Cleopas; but then we also have, in the same list, Zabdon, Zakron, Thorus, Thorisus, Abrazon, etc. What should we do with such a list? To argue that these latter types of names – and there are many more (see our *Book of Bee* dataset in our Supplementary Materials, listed in the references under our names) – are not meant to represent Jewish persons but, rather, a swatch of foreigners sent out by Jesus, alongside his Jewish followers, from Palestine (cf. Luke 10.1), might leave one vulnerable to the same type of circularity that Gregor and Blais accuse Bauckham of when he suggests that the Hellenists in Acts 6.5 (Gregor and Blais 2023: 85) were Diaspora Jews simply because they had Greek names.

We have indicators throughout the *Book of the Bee* that suggest an attempt at onomastic verisimilitude regarding Jewish names, as when a disciple named Ephraim is placed in a list alongside Demas but then also Justus, James, Simon, and John (Budge 1886: 112–13). Demas is similar to Dysmas from *Acts of Pilate*, and Ephraim is a tribal name not dissimilar from fictional

---

[30] The discrepancy possibly comes from the issue discussed in footnote 14 above: Gregor and Blais do not provide an adequate reason for removing 'Adi' ('Ada' under Ilan's orthography) from their *Book of the Bee* sample, so we retain that name; also recall that they incorrectly categorize Elisaeus as 'Eleazar' instead of 'Elisha' (as it should be categorised). Here we want to conduct a robustness check to see if their original sample, without 'Adi' and with 'Elisaeus' as 'Eleazar' instead of 'Elisha', would impact our statistical analysis. The p-value of their sample with these changes is .00004.

Jewish characters like Issachar and Naphtali – also allegedly Jewish persons in *Acts of Pilate* and elsewhere in *Book of the Bee* itself; we have the name Junias (Budge 1886: 111), which appears to be an attempt at the masculine form of the Jewish Palestinian person 'Junia' from Roman 16:7. The list could go on. Ilan appears to cite these other name occurrences in her lexicon rather seldomly and inconsistently. A detailed discussion of this problem by Gregor and Blais would have been fitting.[31] As discussed earlier, Gregor and Blais's data for the *Book of the Bee* is not sufficiently historical to tip the scale in favour of making their apocryphal corpus fit Ilan, so perhaps it is not worthwhile to belabour the point.

# Conclusion

Gregor and Blais (2023) offered the world of New Testament scholarship a valuable contribution with their introduction of statistical methodology to the onomastic congruence insights of Richard Bauckham. However, their methodology was flawed, which Van de Weghe and Wilson (2024) argued led to the wrong conclusions. Unfortunately, they replicated their methodology in their subsequent study of apocryphal and Talmudic corpora. In particular, they employed two different methods: a graphical method assessing the most popular names with 95% confidence intervals and a probability method assessing individual names or clusters of names. Five separate flaws of the graphical method were shown: (i) the wrong conclusion was drawn visibly from their own graph, (ii) the implicit statistical hypothesis test improperly reversed the null and alternative hypotheses, (iii) the small sample size argument fails because it would also say Josephus and other known historical authors were not historical, (iv) the confidence intervals used were for individual names and failed to account for the joint name relationships, and (v) the claim that rare names are "wholly uninformative" regarding distributional fit is false. These flaws cumulatively show that the graphical confidence interval method used in Gregor and Blais (2024) is invalid. The solution is simply to use a standard statistical hypothesis testing method for discrete distributions such as the chi-squared goodness-of-fit test. This was done by Van de Weghe and Wilson (2024) and showed the Gospels-Acts fit the historical data at least as well as Josephus.

---

[31] This discussion is independent from the brief comments made by Gregor and Blais (2024: 285-286) about whether we should follow Ilan's designations regarding occurrences labelled 'fictitious'.

The second method of Gregor and Blais (2024) was the probability method. The four critiques were: (i) the method was used to draw conclusions from rare names even though they previously said rare names were 'wholly uninformative', (ii) the null and alternative hypotheses were reversed from the graphical method, (iii) the method is extremely sensitive to the name assumptions, and (iv) the method fails to address the multiple testing issue. The problem is not that the probability method is invalid – it is valid. The problem is that they used it in contradiction to prior claims, with dubious assumptions, and it is the wrong tool for the job. Therefore, from a statistical perspective, its use gave the impression that they were attempting to use statistics to support a preconceived conclusion. We therefore recommend future onomastic work discontinue those methods and apply proper statistical methodology such as the standard goodness-of-fit test used in Van de Weghe and Wilson (2024).

While carefully assessing various shortcomings in Bauckham's data and observing areas of potential sample bias in *Ilan*, some onomastic features such as qualifiers were overlooked by Gregor and Blais. A holistic and sound conclusion was ultimately lacking in part due to overlooked problems within their own onomastic data. Yet even with the present shortcomings, their own Talmudic and apocryphal corpora have low p-values, running directly contrary to their stated conclusions. In summary, while this latest onomastic study from Gregor and Blais provides some valuable insights, it does not improve on the flaws from their previous article (Gregor and Blais 2023).